\documentclass[conference]{IEEEtran} \IEEEoverridecommandlockouts 
\usepackage{multirow} 
\usepackage{multicol} 
\usepackage{cite} 
\usepackage{amsmath,amssymb,amsfonts} 
\usepackage{algorithmic} 
\usepackage{graphicx} 
\usepackage{textcomp} 
\usepackage{xcolor} 
\def\BibTeX{{\rm B\kern-.05em{\sc i\kern-.025em b}\kern-.08em T\kern-.1667em\lower.7ex\hbox{E}\kern-.125emX}} 

\begin{document} 
\title{A 5G-Edge Architecture for Computational Offloading of Computer Vision Applications} 

\author{\IEEEauthorblockN{1\textsuperscript{st} Marcelo V. B. da Silva}
\IEEEauthorblockA{\textit{Centro de Informática (CIn)} \\ 
\textit{Universidade Federal}\\ 
\textit{de Pernambuco}\\ 
Recife, Brazil \\ 
mvbs3@cin.ufpe.br} 
\and 
\IEEEauthorblockN{2\textsuperscript{nd} Maria Barbosa}
\IEEEauthorblockA{\textit{Centro de Informática (CIn)} \\ 
\textit{Universidade Federal}\\ 
\textit{de Pernambuco}\\ 
Recife, Brazil \\ 
mksb@cin.ufpe.br} 
\and 
\IEEEauthorblockN{3\textsuperscript{rd} Anderson Queiroz}
\IEEEauthorblockA{\textit{Centro de Informática (CIn)} \\ 
\textit{Universidade Federal}\\ 
\textit{de Pernambuco}\\ 
Recife, Brazil \\ 
aalq@cin.ufpe.br} 
\and 
\IEEEauthorblockN{4\textsuperscript{th} Kelvin L. Dias}
\IEEEauthorblockA{\textit{Centro de Informática (CIn)} \\ 
\textit{Universidade Federal}\\ 
\textit{de Pernambuco}\\ 
Recife, Brazil \\ 
kld@cin.ufpe.br} } 

\maketitle

\begin{abstract}
Processing computer vision applications (CVA) on mobile devices is challenging due to limited battery life and computing power. While cloud-based remote processing of CVA offers abundant computational resources, it introduces latency issues that can hinder real-time applications. To overcome this problem, computational offloading to edge servers has been adopted by industry and academic research. Furthermore, 5G access can also benefit CVA with lower latency and higher bandwidth than previous cellular generations. As the number of Mobile Operators and Internet Service providers relying on 5G access is growing, it is of paramount importance to elaborate a solution for supporting real time applications with the assistance of the edge computing. Besides that, open-source based platforms for Multi-access Edge Computing (MEC) and 5G core can be deployed to rapid prototyping and testing applications.  This paper aims at providing an end-to-end solution of open-source MEC and 5G Core platforms along with a commercial 5G Radio. We first conceived a 5G-edge computing environment to assist near to user processing of computer vision applications. Then a sentiment analysis application is developed and integrated to the proposed 5G-Edge architecture. Finally, we conducted a performance evaluation of the proposed solution and compare it against a remote cloud-based approach in order to highlight the benefits of our proposal. The proposed architecture achieved a 260\% throughput performance increase and reduced response time by 71.3\% compared to the remote-cloud-based offloading.

\end{abstract}

\begin{IEEEkeywords}
MEC, Cloud, Offloading, 5G, Edge, Computer Vision.
\end{IEEEkeywords}
\section{Introduction}
\label{sec:intro}
Computer vision applications has gained momentum in various vertical industries and society \cite{CVmarket}. Applications based on object detection, recognition and tracking are starting to be widely deployed in precision agriculture, surveillance, factory floor, smart cities, and education, to cite a few examples. However, those applications require considerable processing power to run machine learning models. From a perspective of end user experience and energy consumption of battery-powered User Equipment (UE), it is unfeasible to rely on smartphones, drones or Internet of Things (IoT) devices, in general, to run those processing hungry applications. To address those applications demands, computer vision processing tasks usually run at remote data centers/cloud computing environments \cite{UAV}.

However, while this method has solved the issues of high power processing and battery consumption in mobile devices, those real time applications also have stringent requirements in terms of latency. Applications that rely on real-time computing require almost instantaneous responses to operate efficiently. The physical location of cloud computing servers directly impacts response delay to applications, which can be problematic in cases where latency is critical, such as in computer vision applications.


Despite the processing benefits of traditional cloud computing, in order to effectively reduce the latency for time-sensitive applications, processing and storage closer to the end-user or data source are required. Nowadays, such an approach is generally described as edge computing, with other terms depending on the access technology, edge location, devices involved, distribution of processing, and context (e.g., fog, mist, and dew computing). By relying on edge computing, time-sensitive applications can offloading their tasks from the mobile device to a more powerful edge server. This allows the mobile device to conserve its battery life and computational resources while leveraging the superior processing capabilities of the edge server. Computation offloading of applications and tasks running on mobile devices to edge servers has been mostly carried out through Wi-Fi or cellular networks (e.g., 4G)\cite{offloading}. 

With the advent of the Fifth Generation (5G) of mobile networks and their widespread deployment by telecom operators in outdoors scenarios and, more recently, as private 5G deployments on premise, real time applications can benefit from the low latency, high bandwidth, and reliability supported by this new generation of cellular networks. But, even with the improvements of 5G air interface over previous generations, keeping the processing at remote cloud might still be harmful for real time applications. Thus, telecommunication operators must also embrace the processing of applications near the end user by adopting edge computing approaches. To this end, the European Telecommunications Standards Institute (ETSI) standardized MEC \cite{ETSI2015}, first coined in 2015 as Mobile Edge Computing and since 2018 as Multi-access Edge Computing.

 The joint 5G-MEC infrastructure seems to be a promising approach since it optimizes response time for sensitive applications and also enables efficient remote processing, providing a more agile and responsive experience for network users. The proposal presented by \cite{UAV} investigates the potential of Unmanned Aerial Vehicles (UAVs) equipped with IoT devices for facial recognition surveillance at high altitudes using offloading video data processing task to MEC servers, however the research is limited to 4G test environments, restricting the exploration of the true potential of low-latency MEC processing in 5G environments. Also, the work  \cite{mar_edge} presents an  exploration of MEC as a solution to mitigate the processing delay and energy consumption inherent in mobile augmented reality applications, the paper uses simulations to evaluate the performance of the proposed hierarchical computation architecture and the results demonstrate that the MEC-based AR framework significantly reduces both energy usage and latency when compared to existing baseline methods, showcasing the effectiveness of the proposed solution for improving mobile AR applications. Based on the state-of-the-art presented, no articles in the current literature were found to consider 5G, computer vision, and MEC/Cloud integrated in a testbed.

As the number of Mobile Operators and Internet Service providers relying on 5G access is growing, it is of paramount importance to elaborate solutions for supporting real time applications with the assistance of edge computing. Besides that, open-source based platforms for MEC and 5G core can be deployed to rapid prototyping and testing applications.  This paper aims at providing a end-to-end solution composed of 5G core, MEC using a commercial 5G radio. We first conceived a 5G-edge computing environment to assist remote (or near to user) processing of computer vision applications. Then a sentiment analysis application is developed and integrated to the proposed 5G-Edge platform. Next, we conducted a performance evaluation of the proposed solution and compare it against a remote cloud-based approach in order to highlight the benefits of our proposal.
\section{Foundation}
\label{sec:funda}
This section provides an overview of basic concepts on computing vision and the building blocks related to the design and implementation of the proposed 5G-MEC infrastructure. First, it is provided an overview of computing vision, followed by the basics of 5G, and finally, the ETSI MEC architecture.

\subsection{Computer Vision}
Computer vision is a field of artificial intelligence that enables computers to extract specific information from images and videos, and from this analysis, various actions can be taken. This technology covers a wide range of algorithms, from traditional image processing techniques to convolutional neural networks (CNNs) and other forms of deep learning \cite{cv_def}.

Computer vision algorithms include object detection, facial recognition, image segmentation, motion tracking, and sentiment recognition, among others. The areas of application are also quite diverse, including healthcare with image-assisted diagnosis and patient monitoring \cite{health}; security with surveillance, access control, and facial recognition systems \cite{UAV}; industry with quality inspections and process automation \cite{industry_fab}; automotive with autonomous cars and driver assistance \cite{cars}.

Computer vision applications are typically executed locally on devices. However, due to high battery consumption \cite{highComputationCost} and limited computing power, there is a shift towards remote processing using cloud servers. While this helps with processing capacity and energy use, it introduces latency challenges that can negatively impact real-time applications. Reducing latency is essential for quick interactions, as delays can disrupt the user experience \cite{Dist5G}\cite{surveyMobileEdge}.

To avoid discomfort, latency in cloud gaming applications should be around 60ms and 100ms for casual gaming, with even lower latencies required for competitive gaming \cite{cloudGaming}. However, for mobile device applications, this minimum latency varies significantly depending on the application, as slightly higher response delays may be sufficient for certain applications.


\subsection{Fifth Generation of Mobile Networks}
The fifth generation of mobile network (5G), offers enhancements on throughput, latency, scalability, and reliability compared to previous generations. It consists of two main components: the Radio Access Network (RAN) and the 5G Core Network (5GC).

The main element of the RAN is the Next Generation Node B (gNodeB), which connects user equipment (UE) to the 5GC. This infrastructure enables wireless communication, supporting data transmission and reception over a broad range of frequencies, including low (sub-1 GHz), mid (1-6 GHz), and high-frequency (mmWave, above 24 GHz) bands, each with unique benefits and challenges.

The 5GC is the central part of the telecommunications infrastructure, divided into the Control Plane (CP) and User Plane (UP). The CP handles functions like authentication, access policies, session establishment, and charging, while the UP, managed by the User Plane Function (UPF), oversees connection management, routing, and data delivery between mobile devices and the internet.

A key innovation in the 5GC is its Service-Based Architecture (SBA), where network functions (NFs) are accessible through standardized APIs, enabling independent implementation, scaling, and updates. This flexibility aligns with the principles of Network Function Virtualization (NFV), which breaks the traditional reliance on Network Equipment Manufacturers by decoupling NFs from hardware. As a result, open-source platforms like OpenAirInterface \footnote{https://openairinterface.org/} (OAI) and Open5GS\footnote{https://open5gs.org/} have emerged as cost-effective solutions, allowing the entire network to be implemented using different platforms, each optimized for specific NFs, access networks, or Edge Computing platforms.

\subsection{ETSI MEC}
The Multi-access Edge Computing (MEC) standard developed by the European Telecommunications Standards Institute (ETSI) extends cloud computing capabilities to the edge of the cellular network. MEC is a system that brings computing, storage, and networking resources closer to users and end devices, reducing latency and improving network efficiency \cite{mecSurvey}. 

\begin{figure}[h!]
    \centering
(ETSI) extends cloud computing capabilities to the edge of    \includegraphics[width=0.48\textwidth]{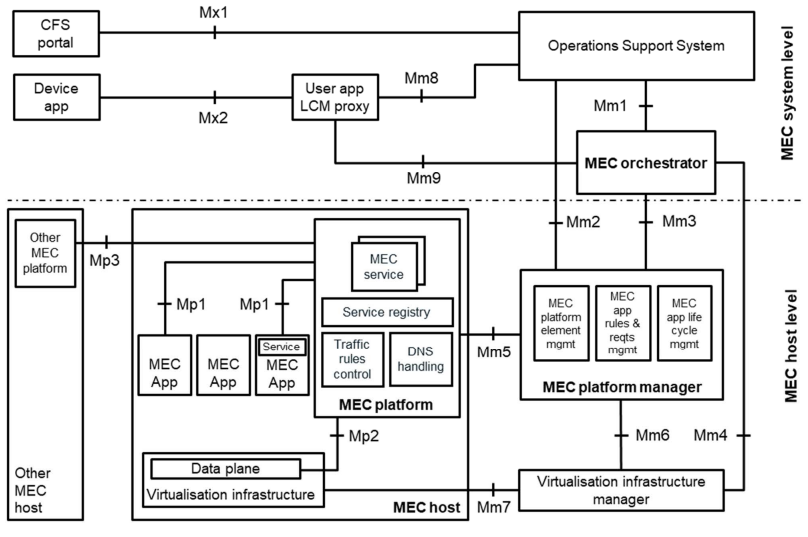}
    \caption{Multi-access edge system reference architecture \cite{ETSI2022}.}
    \label{fig:arcMEc}
\end{figure}

MEC is a key technology for edge computing in 5G, which promises to support new types of applications, ranging from IoT to AR, VR, and autonomous vehicles. The combination of MEC with the 5G infrastructure allows services to be delivered with greater speed and reliability, essential concepts for real-time applications. The standard architecture designed by ETSI is shown in Figure \ref{fig:arcMEc}. It is a relatively complex architecture with many entities and interfaces. For the sake of simplicity, this article specifically focuses on the description of the components deployed in our proposal.


The MEC architecture includes fundamental components such as the MEC platform (MEP), the MEC server, and MEC applications (MEC App):

\begin{itemize}
     \item \textbf{MEC Platform:} The MEP provides the execution environment for edge applications. It manages computing and storage resources, and offers support services such as orchestration and application management. Applications can utilize existing MEC services or register new MEC services. The MEP provides services and their standardized endpoints centrally. 
    \item \textbf{MEC Host} The MEC host, located at strategic points in the network, ensures low latency and high bandwidth by physically hosting the MEC platform.
    \item \textbf{MEC App} Applications developed to run on the MEC Server and utilize their proximity to end users to deliver services with low latency. These applications can have various objectives, essentially functioning as APIs that aim to exploit the advantages of edge location or accessing another MEC service provided by another MEC App. MEC applications must register their services on the MEC platform to be accessible to network users and other MEC Apps. During registration, it is necessary to send a JSON to the MEP specifying all the available services of the application and how to access them.

\end{itemize}

Communication between MEC architectures and the 5GC is facilitated by specific interfaces. The MP1 interface allows a MEC application to communicate with the MEP, while the MP2 interface connects the MEP with the 5GC, it also enables MEC services to access core-related information more directly. The 5GC's data plane serves as the pathway for packets traveling from the UE, through the RAN, to the MEC and cloud platforms. This architecture reduces the number of hops for client application requests to access services provided by MEP compared to cloud-based solutions, resulting in lower latency.

\section{Integrated 5G - MEC Environment for Processing Computer Vision Applications}
\label{sec:props}

\begin{figure}[h]
    \includegraphics[width=0.5\textwidth]{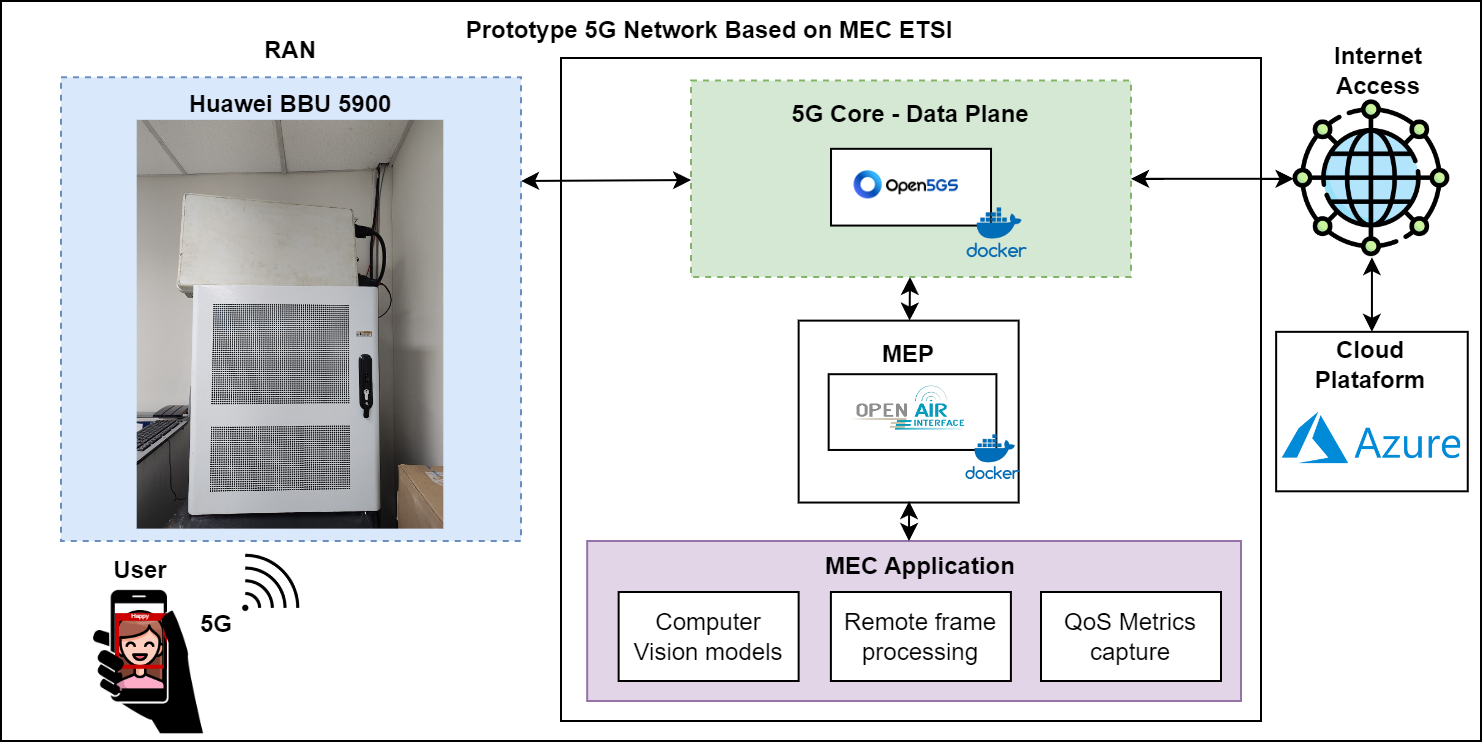}
    \caption{Proposed architecture.}
    \label{fig:arquiteturaUtilizada}
\end{figure}

This section presents the proposed 5G-MEC environment designed to leverage the benefits of low latency for processing computer vision applications. Figure \ref{fig:arquiteturaUtilizada} illustrates the end-to-end architecture. The deployment utilizes open-source platforms for the 5GC and MEC, while the gNodeB and cloud components are vendor-based.

Table  \ref{tab:confenv} summarizes the configurations and components of the deployment. Specifically, Open5GS was used for the 5GC, the ETSI MEC platform was based on OAI, and the gNodeB was a vendor-based solution, namely the Huawei BBU 5900.

\begin{table}[h!]
    \begin{minipage}{\linewidth}
        \centering
        \caption{Environment Configuration.}
        \begin{tabular}{| l | l | l|}
            \hline
            & \textbf{Component} & \textbf{Specification} \\
            \hline
            \multirow{3}{2em}{\textbf{MEC/5GC Host}} & CPU & Intel Xeon Silver 4314 \\ 
            & RAM & 64GB \\ 
            & Operational System & Ubuntu 20.04.6 LTS \\
                \hline
            \multirow{2}{2em}{\textbf{Core}} & Platform & Open5GS \\ 
            & Release 3GPP & Release 17 \\        
            \hline
            \multirow{3}{2em}{\textbf{RAN}} & Vendor & Huawei \\ 
            & Radio & BBU 5900 \\ 
            & Band & n78 \\
            \hline
            \multirow{2}{2em}{\textbf{MEC}} & Platform & OpenAirInterface \\ 
            & Release & ETSI GS MEC 003 V3.1.1 \\ 
            \hline
            \multirow{5}{2em}{\textbf{Cloud Server}} & Platform & 
 Microsoft Azure \\ 
            & Num. vCPU & 2 vCPU \\ 
            &RAM & 8 GB\\
            &Operational System & Ubuntu 20.04.6 LTS \\ 
            &Localization & Brazil South (Zone 3) \\ 
            \hline
            \multirow{2}{4em}{\textbf{User}} & User Terminals & Motorola Edge 30 Ultra/Neo \\
            & Sim Card & Sysmocom - S1J1. \\
            \hline
        \end{tabular}
        \label{tab:confenv}
    \end{minipage}
\end{table}

 The MEC application, developed in Flask, was registered on the MEP upon initialization, making all its endpoints available to users.  

\begin{figure}[h!]
    \centering
    \includegraphics[width=0.5\textwidth]{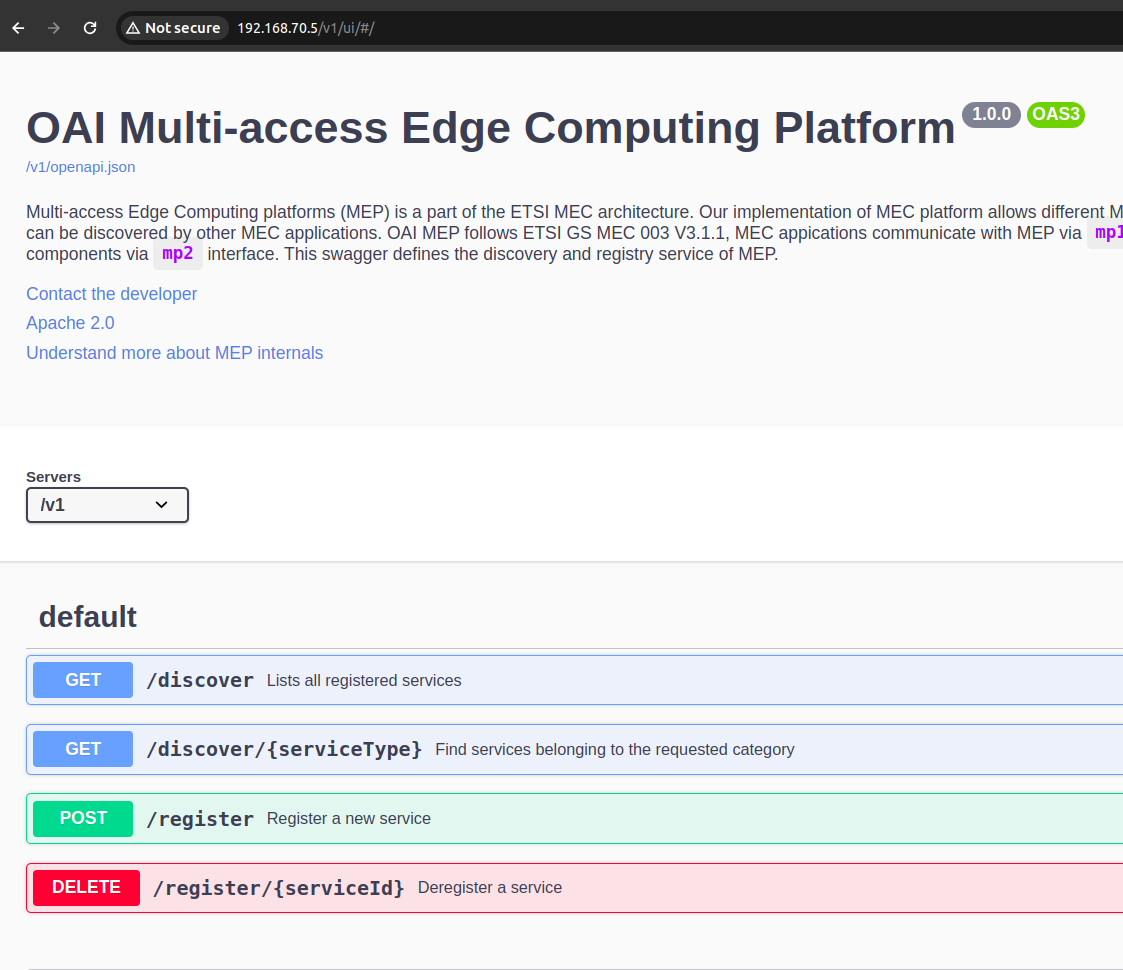}
    \caption{MEC Platform swagger and endpoints.}
    \label{fig:swagger}
\end{figure}

The Figure \ref{fig:swagger} shows the graphical interface of the MEP API that centralizes MEC services. This API interface is called Swagger and includes all the MEC API endpoints. As depicted in the Figure, the developers can visualize the commands to register and unregister MEC services, as well as the command to discover all available MEC services. HTTP requests can be executed through the browser or via any standard HTTP request as well.


Finally, the Cloud platform was implemented and hosted on the Microsoft Azure. The cloud application has the same functionalities and endpoints as the MEC application, with the only difference being the manner of deployment.



\section{Sentiment Analysis Application}
\label{sec:app}

In order to demonstrate the effectiveness of the proposed 5G-Edge environment for offloading  computer applications, a implementation of sentiment analysis application was carried out as a use case. This application aims to detect and classify human emotions from real-time video images, with remote processing done both at the edge network by the MEC app and in other scenario in a remote server done by the cloud app.

The Android application was developed in React Native and captures each frame from the camera, sending them via HTTP requests to remote servers. These servers process the images using sentiment recognition algorithms and return the face location along with the detected sentiment. Based on this return, the Android application displays all detected faces and the sentiment of each detected face on the screen.

For these remote processing applications, a Flask\footnote{https://flask.palletsprojects.com/}
 server in Python\footnote{https://www.python.org/} was used, employing the FER library. This library uses the Multi-task Cascaded Convolutional Networks (MTCNN) technique for face detection and a pre-trained model for classifying sentiments into six categories: ‘fear’, ‘neutral’, ‘happy’, ‘sad’, ‘anger’, and ‘disgust’. The choice of this algorithm is motivated by its relevance in current researches and practical applications, such as measuring interest in educational environments and therapies for children with Autism Spectrum Disorder (ASD)\cite{tea}.

The servers also capture various metrics while running to evaluate the performance of the computer vision application on the remote cloud or MEC contexts.
\section{Performance Evaluation}

To assess the effectiveness and performance of the proposed computer vision application, two devices were used: the Motorola Edge 30 Ultra and the Motorola Edge 30 Neo. The application offers two processing options: cloud processing and MEC (Multi-access Edge Computing) processing. Both smartphones were connected to the 5G network. Measurements were taken using each phone individually, as well as both phones making requests simultaneously.

The testing methodology involved using the above-mentioned application. During each request, the current frame from the camera was sent with a resolution of 200x152 pixels. Resizing the frame was necessary to speed up the availability of the frame for the phone and to reduce the size of the image sent over the network, which also decreases processing time on the cloud and MEC servers.

The scenarios were evaluated based on the following metrics:

\begin{itemize}
\item \textbf{RTT (Round-Trip Time)}: It is the latency, measured in milliseconds, a packet takes from the phone to the server and back to the phone. 
\item \textbf{sentiment Recognition Algorithm Processing Time}: This time refers to the period required for the remote server to process the received image and determine the sentiment present in the frame and the location of faces on the screen. The metric was measured in millisecond.
\item \textbf{Response Time}: This is the time from sending the image from the phone to receiving the server's response on the user end. This metric is crucial for understanding the total latency involved in the sentiment recognition process. The metric was measured in millisecond.
\item \textbf{Throughput}: Throughput was calculated by dividing the size of the data sent and received by the response time. This metric is important for evaluating the network's efficiency in terms of the volume of data transmitted per unit of time. The metric was calculated in Megabits per second.
\end{itemize}

The results obtained from these metrics allow for a detailed analysis of the application's performance under different processing and connectivity conditions, highlighting the advantages and limitations of using cloud and MEC for computer vision applications on 5G-connected mobile devices.

\subsection{Results}
This section presents the results of the performance comparison between the remote cloud approach and the MEC-based solution within the 5G end-to-end architecture.

In the remote cloud configuration, “Cloud1” refers to the scenario utilizing one device, while “Cloud2” involves two devices. Similarly, in the MEC-based approach, “MEC1” and “MEC2” correspond to setups with one and two devices, respectively. To ensure a 95\% confidence interval, one hundred samples were collected for each scenario.

\subsubsection{RTT}
\begin{figure} [h!]
    \includegraphics[width=0.45\textwidth]{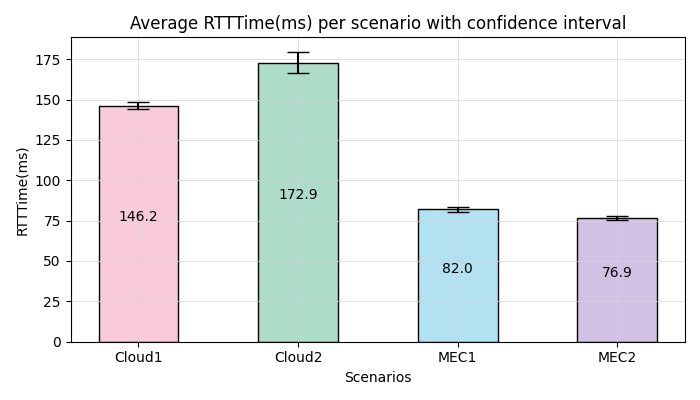}
    \caption{Average RTT time to MEC and Cloud scenarios measured to one and two devices.}
    \label{fig:rttMedio}
\end{figure}

As shown in Figure \ref{fig:rttMedio}, the average RTT (Round-Trip Time) was substantially lower in scenarios where processing was performed in MEC (MEC1 and MEC2) compared to cloud-based processing (Cloud1 and Cloud2). The lower RTT observed in MEC (82.0 ms and 76.9 ms) reflects the advantage of processing data closer to the end-user, this way reducing network latency. This result confirms the expectation that MEC can provide a quicker response due to its proximity to mobile devices, which is crucial for computer vision applications requiring real-time processing.

\subsubsection{Processing Time}
\begin{figure}[h!]
    \includegraphics[width=0.45\textwidth]{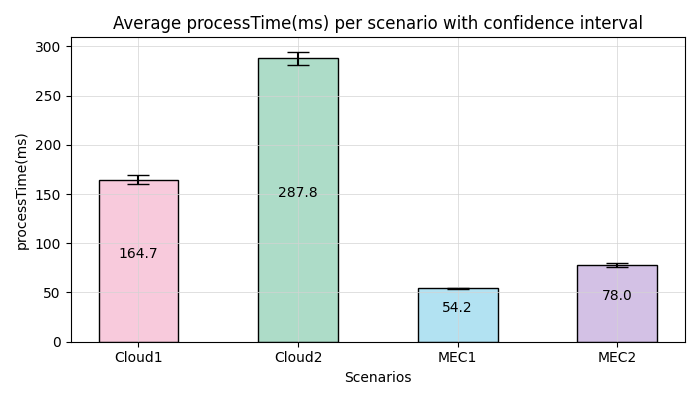}
    \caption{Average processing time to MEC and Cloud scenarios measured to one and two devices.}
    \label{fig:processTimeMedio}
\end{figure}

The processing times of the sentiment recognition algorithm, presented in Figure \ref{fig:processTimeMedio}, demonstrate that MEC is significantly more efficient (54.2 ms and 78.0 ms) compared to the cloud (164.7 ms and 287.8 ms).This translates to remarkable improvements of 67.5\% and 73.1\% for MEC compared to cloud computing in each scenario respectively. The advantage of MEC can be attributed to lower network overhead and the ability to allocate computational resources more effectively, including the computational power of the cloud machine. These results are due to the limited processing hardware used on the cloud platform, where this limitation directly impacts scenarios with more connected devices because greater computational power is needed.

\subsubsection{Response time}
\begin{figure}[h!]
    \includegraphics[width=0.45\textwidth]{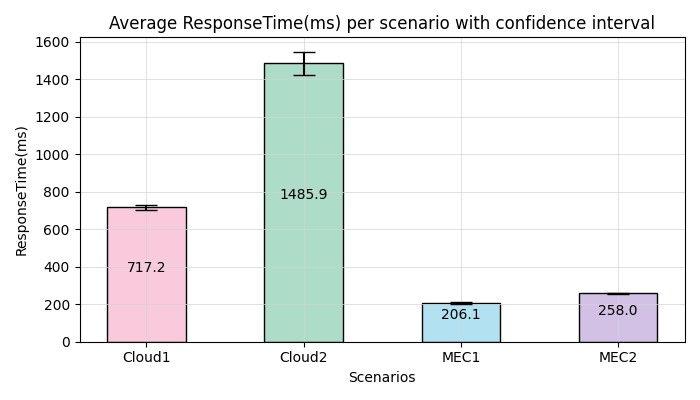}
    \caption{Average response time  time to MEC and Cloud scenarios measured to one and two devices.}
    \label{fig:ResponseTimeMedio}
\end{figure}

The total response time, which includes both RTT and processing time, also favors MEC, as shown in Figure \ref{fig:ResponseTimeMedio}. With response times of 206.1 ms and 258.0 ms for MEC, compared to 717.2 ms and 1483.9 ms for the cloud, it is clear that MEC offers superior performance. Specifically, MEC shows a performance improvement of approximately 71.3\% and 82.6\% over the cloud. This metric is particularly important for computer vision applications, where noticeable delays can compromise the functionality and usefulness of the application.

\begin{figure}[ht!]
\centering
\includegraphics[width=0.35\textwidth]{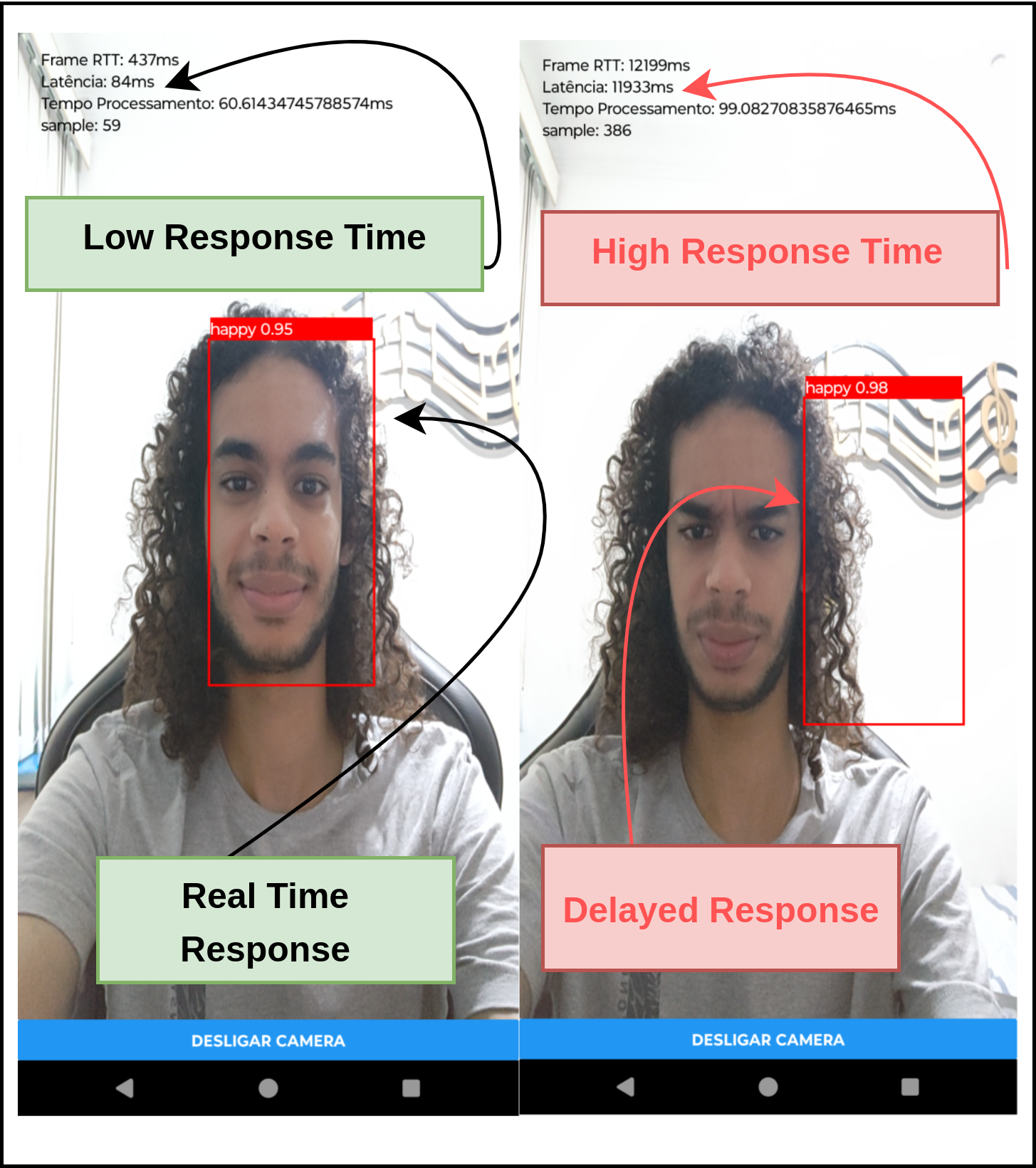}
    \caption{Latency impact to a computer vision app.}
    \label{fig:ImpactoLatencia}
\end{figure}

The impacts of high response time can be observed in Figure \ref{fig:ImpactoLatencia}. A high response time results in delays in face and sentiment detection, which can lead to errors as an incorrect location on detection. 

\subsubsection{Throughput}

\begin{figure}[h!]
    \includegraphics[width=0.45\textwidth]{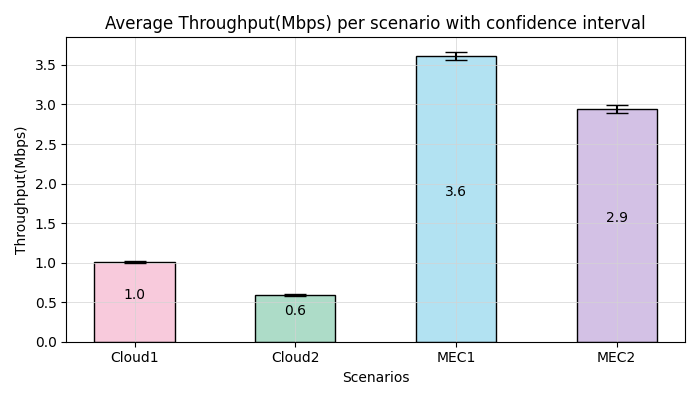}
    \caption{Average throughput to MEC and Cloud scenarios measured to one and two devices.}
    \label{fig:thoughputMedio}
\end{figure}

Finally, Figure \ref{fig:thoughputMedio} shows that the average throughput was significantly higher in MEC scenarios (3.6 Mbps and 2.9 Mbps) compared to the cloud (1.0 Mbps and 0.6 Mbps). MEC shows a performance improvement of approximately 260\% and 383.33\% over the cloud. This increase in throughput indicates that MEC can transmit and process larger volumes of data more quickly, making better use of available bandwidth. This result is crucial for computer vision applications that rely on continuous and high-quality video streaming, as higher throughput allows more data to be transmitted in less time, improving overall system efficiency.

\label{sec:resu}

\section{Conclusion}


The joint approach 5G-MEC can reduce the latency and enhance the efficiency of these applications. Multi-access Edge Computing allows processing and storage closer to the end user, while 5G provides a network infrastructure with high speed, low latency, and increased capacity for simultaneous device connections. This paper proposed and deployed a 5G-MEC architecture to support computer vision applications, utilizing a combination of open-source software (5G Core and MEC) and a commercial 5G radio.

The results obtained consistently demonstrate that the use of MEC alongside 5G networks delivers substantial latency and throughput gains for computer vision applications. MEC reduces the latency, making it a superior solution to cloud computing for applications demanding real-time processing with low response time. While processing time metrics are also influenced by the computational power of edge and cloud servers, the results reveal that the most significant performance difference happens because data transfer time between the device and the server. Due to its proximity to devices and reduced data transfer requirements, MEC offers significantly lower transmission time compared to cloud computing.

For future work, it is essential to expand the number of users and analyze the consumption of computational resources and energy. Additionally, efforts should focus on enhancing the application to improve the transmission rate.

\label{sec:conc}

\section*{Acknowledgment}
This work was supported by the National Council for Scientifc and Technological Development (CNPq) - Research Productivity Fellowship (Grant No. 313083/2023-1), Pernambuco Research Foundation (FACEPE) (Grant No. IBPG 1602-1.03/24), National Institute of Science and Technology for Software Engineering (INES), and Telecom Italia (TIM)


\end{document}